# Effect of molecular and electronic structure on the light harvesting properties of dye sensitizers

*E. Mete\*, D. Uner, M. Çakmak, O. Gulseren, and Ş. Ellialtıoğlu*

Department of Physics, Balikesir University, Balikesir 10145, Turkey

Department of Chem. Eng., Middle East Technical University, Ankara 06531, Turkey

Department of Physics, Gazi University, Ankara 06500, Turkey

Department of Physics, Bilkent University, Ankara 06800, Turkey

Department of Physics, Middle East Technical University, Ankara 06531, Turkey

ABSTRACT

The systematic trends in structural and electronic properties of perylene diimide (PDI) derived dye molecules have been investigated by DFT calculations based on projector augmented wave (PAW) method including gradient corrected exchange–correlation effects. TDDFT calculations have been performed to study the visible absorbance activity of these complexes. The effect of different ligands and halogen atoms attached to PDI were studied to characterize the light harvesting properties. The atomic size and electronegativity of the halogen were observed to alter the relaxed molecular geometries which in turn influenced the electronic behavior of the dye molecules. Ground state molecular structure of isolated dye molecules studied in this work depends on both the halogen atom and the carboxylic acid groups. DFT calculations revealed that the carboxylic acid ligands did not play an important role in changing the HOMO–LUMO gap of the sensitizer. However, they serve as anchor between the PDI and substrate $TiO_2$ surface of the solar cell or photocatalyst. A commercially available dye-sensitizer, ruthenium bipyridine $[Ru(bpy)_3]^{2+}$ (RuBpy), was also studied for electronic and structural properties in order to make a comparison with PDI derivatives for light harvesting properties. Results of this work suggest that fluorinated, chlorinated, brominated, and iyodinated PDI compounds can be useful as sensitizers in solar cells and in artificial photosynthesis.

INTRODUCTION

The growing demand on energy and increasing levels of atmospheric carbon dioxide as a result of combustion related energy generation processes have driven the research towards direct or indirect solar based energy generation. Direct utilization of solar energy is possible through photovoltaics or dye sensitized solar cells (DSSC) where absorbed photons are used to generate electrons. Indirect utilization also requires absorption of photons to generate electrons, however, in this case the electrons are used in chemical conversion, namely, artificial photosynthesis. The needs for developing materials for such systems are several fold and listed in detail in a recent article[1]. Briefly summarized they can be grouped into three:

*\* E-mail :* emete@balikesir.edu.tr  *Tel :* +90 (266) 612 1000  *Fax :* +90 (266) 612 1215

(i) the development of new semiconductor nano-crystals containing homo and hetero p–n junctions for efficient light harvesting and charge separation
(ii) the development of efficient catalytic systems capable of utilizing photogenerated electrons in further chemical conversions
(iii) the development of efficient organic or inorganic light harvesting systems, capable of generating electrons and donating these electrons to the superstructure for electron utilization (in photovoltaics or in DSSCs) or for further chemical conversions (in artificial photosynthesis).

Absorption of light with subsequent generation of electrons can be accomplished by the use of dye sensitizers. In such systems a donor and an acceptor are connected to a chromophore which is capable of generating an electron as a result of photon excitation[2]. Such systems can be used in solution to initiate chemical reactions, or they can be incorporated into complex frameworks including electrodes and electrolytes for photoelectrochemical synthesis (PES)[3], or for dye sensitized solar cells[4]. The relative ease of tailoring the dye sensitizers for the desired wavelength photon absorption makes them an attractive solution for photon utilization problems. Furthermore, as more is learned from natural photosynthetic systems, better mimicking can be possible.

Dye sensitizers constitute a vast area of current research. Their synthesis is a specialized field requiring expensive chemicals and procedures. On the other hand, the diversity with which those synthesizers can be produced makes them enormously attractive. The ultimately tailored dye sensitizer should be computer generated with all the physical and chemical properties predicted in advance and synthesized subsequently. It is possible to obtain the *ab initio* electronic structure[5] and therefore to predict the absorption and emission spectra of the dyes[6], to systematically design and study the materials from dyes[7, 8, 9], and to molecularly design dyes for DSSC, based upon the predictions on the effects of the functional groups and on the electronic levels[10]. It is also possible to predict the interactions of the dyes with the substrate, most notably $TiO_2$[11, 12].

Ruthenium based dye complexes are considered for solar cell devices due to their high efficiency in sensitizing $TiO_2$ nanocrystalline particles[13]. These rare earth metal dyes prove to be successful in such applications because of their excellent light harvesting electronic spectra[14,15], fast charge injection kinetics[16,17] and long-term molecular stability under various environmental conditions[18,19,20]. Their electronic structures have been studied extensively[16,17,21,22,23]. In addition to these experimental works there are several theoretical studies[24,25,26], as well, which focus on the electronic spectra of these chromophores.

On the other hand, finding alternatives to these expensive metal driven dye sensitizers presently raised considerable research interest[27,28]. Recently, an organic dye molecule, indoline, has been shown to be highly efficient in light harvesting for DSSC applications[28]. Among the other candidates, the perylene derivatives with carboxylic acid groups suggested by Ferrere *et al.*[27] have been found to be successful in sensitizing dye–semiconductor systems. Perylene dyes are highly absorbing and emitting in the visible range with quantum yields near unity[29]. They have excellent heat and chemical stability with lightfast electron transfer capability from dye singlet state to semiconductor conduction band. In addition to these good properties, they allow to be engineered by attaching different ligands for altering physical and chemical properties.

Lately, perylenediimide (PDI) based dye compounds have been considered for photodynamic therapy[30] and in artificial photosynthesis[31]. The latter work studied brominated PDI derivatives, BrGly and BrAsp, which were synthesized for testing the gas phase $CO_2$ photoreduction efficiencies on $TiO_2$ film surfaces. We report the molecular and electronic structure of these dye compounds and commercially available $[Ru(Bpy)_3]^{2+}$ (RuBpy) dye complex, for comparison. The chemical schematics for BrGly, BrAsp and RuBpy as well as other possible PDI derivatives are presented in Figure 1. In addition to the determination of molecular structures of these dyes, *ab initio* calculations have been performed to study the effect of functional carboxylic groups, $R_1$, and of halogen atom, $R_2$, on the molecular and electronic structures of PDI derivatives.

THEORETICAL CALCULATIONS

DFT calculations have been performed with the VASP[32, 33] code. The electron–ion interactions have been described by projector-augmented wave (PAW) method[34,35]. The wavefunctions were represented in terms of plane-wave expansion. The gradient corrected exchange and correlation effects were taken into consideration using Perdew–Burke–Ernzerhof (PBE) scheme with Perdew's Pade approximation for the LDA part[36].

Plane waves with kinetic energies up to the cut-off value of 29.4 Ryd were included in the basis sets, in order to expand the wavefunctions for all the molecules: PDI-based dyes and RuBpy. For the geometry optimizations, the dye molecules were placed in orthorhombic supercells which contain at least 8 Å vacuum regions in all 3 cartesian directions in order to ensure that no charge transfer occurs between the molecular edges. These isolated structures were relaxed into their minimum energy configurations using the conjugate gradients algorithm by requiring the forces on each atom to be smaller than 10 meV/Å. All calculations were done at the Γ point since we have large supercells which prevent any dispersion of molecular states in the BZ. In fact, a 4 k-point BZ integration shows no significant change in the total energies up to a tolerance of $10^{-5}$ eV. In this sense, we present very well converged results which manifest relaxed electronic charge distributions as well as minimized stresses and forces on atoms.

In order to find the relaxed geometric structure which corresponds to the ground state minima of the BrGly and BrAsp dyes, we started with the perylenediimide which constitutes a base for both of these molecules. After obtaining the atomically relaxed arrangement of perylenediimide, we added the required carboxyl groups accordingly and re-relaxed to get the minima of potential energy surfaces (PES) of BrGly and BrAsp agents. This procedure is also useful in understanding the electronic structure of these dye molecules which are derived from the basic molecular structure of PDI. RuBpy, on the other hand, has been relaxed from a roughly estimated initial configuration by imposing the same tolerance value as used for the other two dye complexes.

We have considered PDI with different halogen ions, namely F, Cl, Br, and I, bonded to it as well as various carboxyl groups attached to it, namely the glycine (Gly) and aspartine (Asp) groups. We have started our calculations by relaxing the bare PDI to its minimum energy configuration. Later on we have used this relaxed geometry for the starting point in calculating the relaxed geometries of the other eight related dye complexes. We have also studied the structural and electronic properties of RuBpy in order to bring a fundamental understanding to the characteristics of light absorbing and charge injecting states from a physical standpoint. In addition to the energy levels and the HOMO–LUMO differences, we have also calculated the twist angles of the PDI skeleton in each dye molecule, as well as various bond lengths and binding energies which are tabulated for comparison. The effects of halogen ions and of additional ligands on the electronic properties, and on their possible role in catalytic properties are discussed.

In order to investigate the absorbance characteristics of these dyes in vacuo, we performed time dependent density functional theory (TDDFT) calculations using NWChem[37,38] package with the Becke three-parameter-Lee-Yang-Parr (B3LYP) hybrid–DFT exchange-correlation functional[39,40]. Standard Gaussian basis functions 6–31G* and 6–311G* were employed for organic species (H,C,N,O) and for halogen group (F,Cl,Br,I), respectively. On the other hand, we used relativistic effective core potential for Ruthenium with the LANL2DZ basis set[41]. We utilized plane wave GGA–PBE96 results as the starting configuration for DFT-B3LYP geometry optimizations. The mixed functional B3LYP is parametrized to yield excellent agreement with experimental results and is good for nonperiodic systems. Plane wave based PBE96 functional is known to reproduce bond lengths reasonably accurate and is suited for extended systems such as molecule-TiO$_2$ interfaces. We have seen that the self-consistent geometry optimizations converged in a few iterations. Hence, the difference between the molecular structures obtained with B3LYP and PBE96 calculations are insignificantly small. Then, TDDFT calculations were carried out on top of these hybrid–DFT re-optimized structures. Since the exchange-correlation potentials generated from B3LYP functional decay as the true Coulomb potential, and since a few low lying vertical excitations are considered in this study, TDDFT methods produce accurate results for describing visible range absoption spectra of these dye complexes.

RESULTS AND DISCUSSION

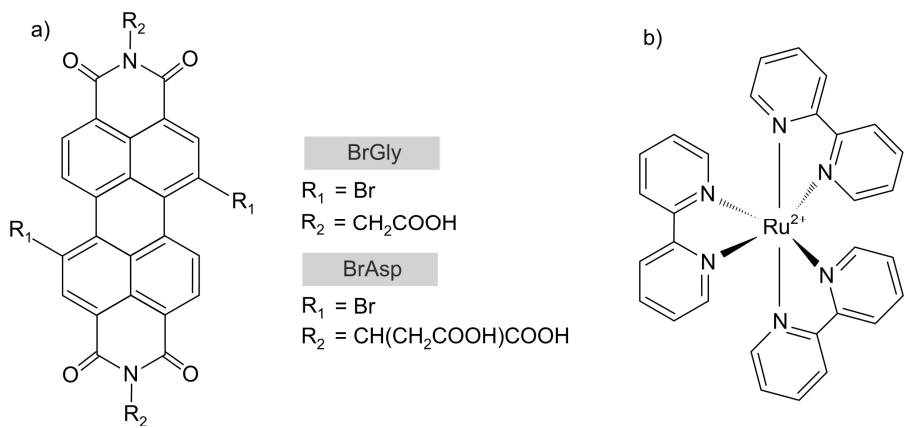

**Figure 1.** Chemical drawings for (a) PDI based, and (b) RuBpy dye molecules.

*A priori* knowledge of the electronic properties of the dye sensitizers is very important in designing efficient systems that utilize solar energy, such as dye sensitized solar cells (DSSC)[4]. In order to bring about a quantum mechanical understanding to the molecular structure and photocatalytic properties of PDI-derived dye sensitizers we performed first principles DFT calculations. The band gaps, HOMO and LUMO levels of the sensitizers were determined as described in the previous section. In this study, we selected PDI derivatives with proven photodynamic therapy[30] and artificial photosynthesis[31] properties. We extended the work to observe the effect of the halogen atom on the molecular and electronic structure of the PDI derivatives. Furthermore, we determined the molecular and electronic structure of commercially available [Ru(Bpy)$_3$]$^{2+}$. A stepwise approach has been chosen for the calculations. First, we took the PDI core and relaxed the structure (Fig 2.a). On the relaxed structure first the halogen atom (Fig 2.b) and then carboxylic acid groups (Fig 2, inset, top left) were attached and these structures were relaxed. Minimum energy structures of BrGly and BrAsp molecules are shown in Figure 2.c and Figure 2.d, respectively, for reference. Corresponding molecular energy levels of the dyes, which might have contributions to the lowest lying optical excitations, are also shown in Figure 2.

It is well known that LDA underestimates the magnitude of the band gaps. Hence, the DFT energy gap values for PDI, Br-PDI, BrGly, and BrAsp, which are presented in Table 1, show slightly infrared characteristics. These dye molecules exhibit luminescence near green region[30] which corresponds to an underestimation of magnitude ~0.8 eV. On the other hand, TDDFT corrected gaps which are assigned to HOMO-LUMO transitions reproduce experimental results. Similarly, the gap width is predicted to be 1.85 eV and 2.64 eV by DFT and TDDFT calculations, respectively, while the experimental value is around 2.7 eV depending on the chemical environment[42]. In order to study systematic trends in calculated values for PDI-derived dye molecules, we included fluorinated, chlorinated, and iodinated PDI (F-PDI, Cl-PDI, and I-PDI, respectively), however, to the extend of the authors' knowledge, there is no experimental work on these molecules. Therefore, we also present the results for these PDI species on Table 2 to elucidate the role of the halogen atom on the band gap. It was observed that when Br atoms were replaced by F atoms, the twist angle was decreased from 27.9° to 15.1° due to the relatively more localized electron density around F ions leading to relatively less repulsive interaction with the nearest-neighbor H atom. Another factor might be the fact that F atom forms a closed shell system electronically when it binds to the PDI. Furthermore, F-PDI has exhibited a band gap which is 50 meV (DFT) and 110 meV (TDDFT) larger than the band gap of Br-PDI.

**Table 1.** Calculated values of typical energies (eV), various bond lengths (Å), and twist angles for the relevant dye molecules.

| Complex | $\Delta E_{HL}$ TDDFT | $\Delta E_{HL}$ DFT | $d_{X-C}$ [a] | $d_{PDI-cg}$ [b] | $d_{Ru-N(bpy)}$ [c] | $d_{C-C}$ | $d_{N-C}$ | $d_{O-C}$ | $BE_{cg}$ [d] | $\theta$ (°) [e] |
|---|---|---|---|---|---|---|---|---|---|---|
| RuBpy | 2.64 | 1.85 | – | – | 2.07 | 1.40 | 1.37 | – | – | – |
| PDI | 2.45 | 1.47 | – | – | – | 1.41 | 1.38 | 1.22 | – | 7.6 |
| Br-PDI | 2.39 | 1.45 | 1.88 | – | – | 1.41 | 1.38 | 1.22 | – | 27.9 |
| BrGly | 2.37 | 1.45 | 1.89 | 1.45 | – | 1.43 | 1.42 | 1.26 | 4.80 | 27.7 |
| BrAsp | 2.36 | 1.44 | 1.89 | 1.47 | – | 1.44 | 1.42 | 1.27 | 4.36 | 28.2 |

[a] $X$ refers to the corresponding halogen element.
[b] Bond length between the carboxyl groups (cg) and the PDI skeleton.
[c] Ru-N bond distance.
[d] Binding energy of attached carboxyl groups (cg).
[e] Twist angle from the planar geometry.

**Table 2.** The effect of the halogen on the typical energies (eV), various bond lengths (Å), and twist angles of PDI derived dye molecules.

| Chromophore | $\Delta E_{HL}$ TDDFT | $\Delta E_{HL}$ DFT | $d_{X-C}$ [a] | $d_{PDI-cg}$ [b] | $d_{C-C}$ | $d_{N-C}$ | $d_{O-C}$ | $BE_{cg}$ [c] | $\theta$ (°) [d] |
|---|---|---|---|---|---|---|---|---|---|
| PDI | 2.45 | 1.47 | – | – | 1.41 | 1.38 | 1.22 | – | 7.6 |
| F-PDI | 2.50 | 1.50 | 1.38 | – | 1.43 | 1.39 | 1.23 | – | 15.1 |
| FGly | 2.48 | 1.49 | 1.36 | 1.45 | 1.43 | 1.42 | 1.26 | 4.85 | 14.9 |
| Fasp | 2.46 | 1.49 | 1.36 | 1.47 | 1.44 | 1.42 | 1.26 | 4.42 | 15.9 |
| Cl-PDI | 2.44 | 1.46 | 1.74 | – | 1.42 | 1.40 | 1.23 | – | 27.9 |
| ClGly | 2.43 | 1.46 | 1.74 | 1.45 | 1.43 | 1.42 | 1.26 | 4.85 | 27.3 |
| ClAsp | 2.41 | 1.45 | 1.74 | 1.47 | 1.44 | 1.43 | 1.27 | 4.41 | 28.0 |
| Br-PDI | 2.39 | 1.45 | 1.88 | – | 1.42 | 1.38 | 1.22 | – | 27.9 |
| BrGly | 2.37 | 1.45 | 1.89 | 1.45 | 1.43 | 1.42 | 1.26 | 4.80 | 27.7 |
| BrAsp | 2.36 | 1.44 | 1.89 | 1.47 | 1.44 | 1.42 | 1.27 | 4.36 | 28.2 |
| I-PDI | 2.31 | 1.40 | 2.12 | – | 1.42 | 1.40 | 1.23 | – | 28.2 |
| IGly | 2.30 | 1.40 | 2.12 | 1.45 | 1.43 | 1.42 | 1.26 | 4.82 | 28.1 |
| IAsp | 2.28 | 1.39 | 2.12 | 1.47 | 1.44 | 1.43 | 1.27 | 4.38 | 29.6 |

[a] $X$ refers to the corresponding halogen element.
[b] Bond length between the carboxyl groups (cg) and the PDI skeleton.
[c] Binding energy of attached carboxyl groups (cg)
[d] Twist angle from the planar geometry

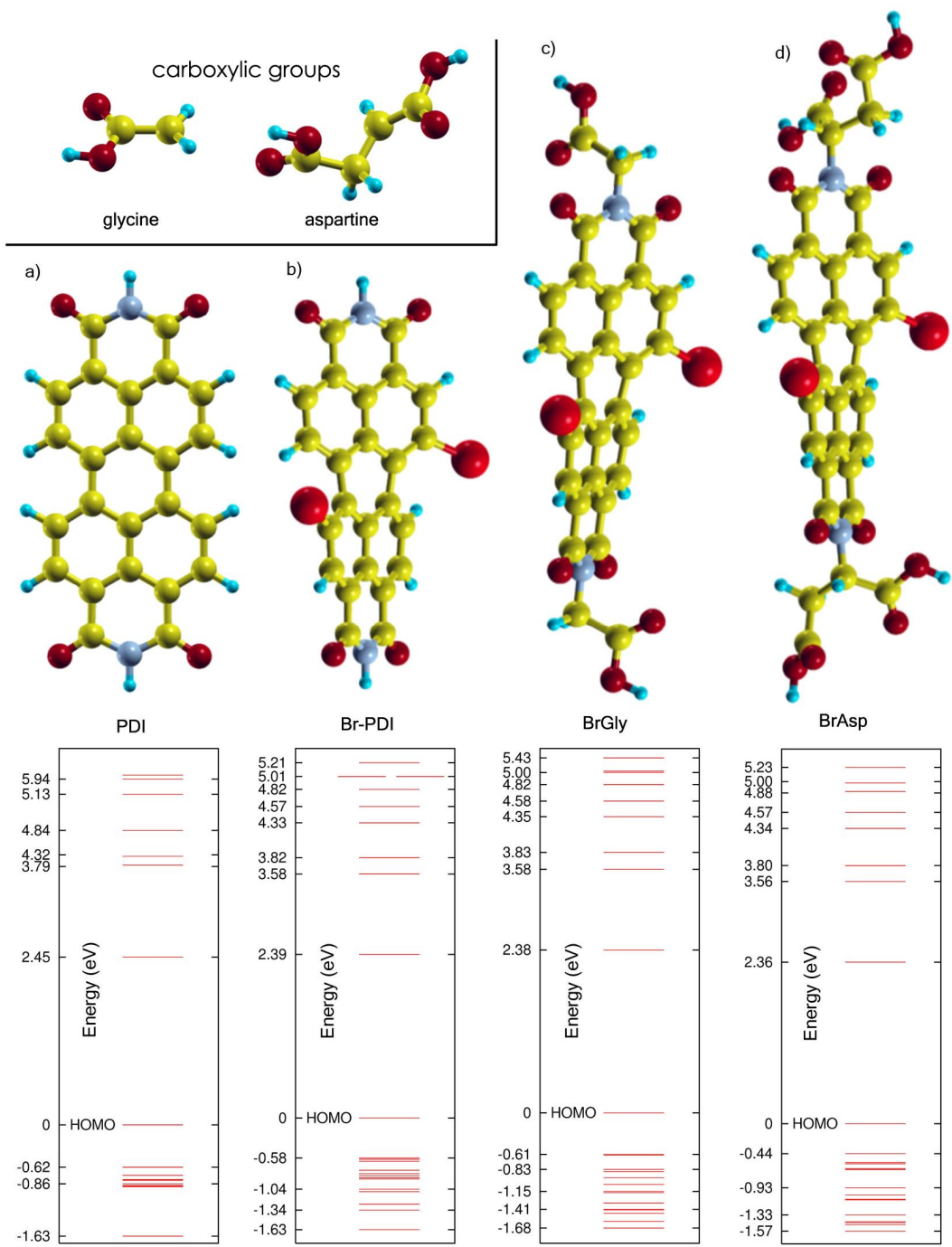

**Figure 2.** Calculated molecular structures and TDDFT corrected molecular orbital levels of PDI based dye molecules.

We observed a local minimum for Br-PDI structure in which the perylene part is planar. This transition state has a HOMO–LUMO Kohn-Sham gap of 1.57 eV. On the other hand, the minimum energy configuration gives a gap of 1.45 eV. Between these two relaxed geometries the only difference is that the perylene skeleton is slightly twisted due to the largely spread wavefunctions of Br atoms, in the latter case. The twisting of the molecule might be a result of the repulsive stress between Br and its nearest neighbor hydrogen. There is a difference of 0.44 eV in the total energies between these two cases. We observed that the gap got narrower by 0.12 eV and the HOMO level shifted down by 0.09 eV from planar to twisted. When the Br atoms are replaced with hydrogen atoms, the gap values became 1.53 eV and 1.47 eV for the planar and twisted geometries, respectively. Thus, for perylene construction, the HOMO–LUMO energy gap depends on two main factors: (1) the choice of the halogen species, and (2) the low energy geometric structure of the molecule.

When we consider more complex clusters like BrGly and BrAsp, another factor emanates, which is the effect of ligands that are attached to the Br-PDI. BrGly and BrAsp have TDDFT corrected HOMO–LUMO gaps of 2.37 eV and 2.36 eV, respectively. Although the difference in gap sizes is small, these two complexes show different reactivity profiles due to the attached parts. The relaxed geometries of these ligands are shown in Figure 2(inset). A significant reconstruction is observed when they bind to Br-PDI to form BrGly or BrAsp structures which are shown in Figure 2c and 2d, respectively. This can be considered as a result of the electrostatic field which derives them to their lowest energy states. Thus, for both of these dyes, the occurrence of the maximum C–C bond length at the tips within $CH_2COOH$ and $CH(CH_2COOH)COOH$ groups rather than around the twist center, might be explained within this context.

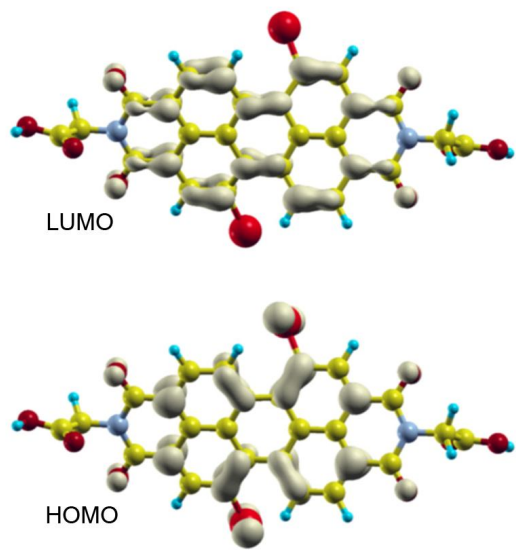

**Figure 3.** Calculated HOMO and LUMO charge density distributions of BrGly molecule.

In addition, we have found that these ligands do not contribute to the HOMO and LUMO, electronically. For instance, one can readily see in Figure 3 that $CH_2COOH$ group has no charge density distribution, neither on the HOMO nor on the LUMO levels of BrGly. This holds for the case of BrAsp, too. Their effect, rather, is seen in the molecular orbitals up to no more than three energy levels (or ~0.84 eV) below the HOMO. Even though the attached carboxyl groups have no electronic contribution in the HOMO and LUMO, one cannot say that their effect in the energy gap is none. Data in Table 2 show the HOMO–LUMO gaps of Br-PDI, BrGly, BrAsp, and other PDI-based dye molecules. These values are slightly different from each other and exhibit a correlation with the twist angle around the major axis which passes from one tip to another through the molecule. As the twist angle of relaxed molecular structure increases, the HOMO–LUMO energy gap gets narrower. This is the case when comparing the gap values for relaxed geometries of different halogenated PDI derivatives. On the other hand, when each molecule is twisted further, the gap value start to increase for that molecule as the geometry goes away from the relaxed structure. As an example, we present the results for PDI and Br-PDI in Figure 4. We started with unrelaxed planar structures for both of the molecules and slightly twist them at each step of the calculation. We did not relax these structures but we calculated the HOMO–LUMO gaps for each twist angle chosen. The findings for these roughly estimated intermediate molecular shapes imply that the minimum gap value occurs at the geometry which corresponds to the relaxed structure. Therefore, our results suggest that the energy gap value depends dominantly on the geometry of the perylene part and the choice of the halogen atom, rather than on the electronic contribution by the carboxyl ligands.

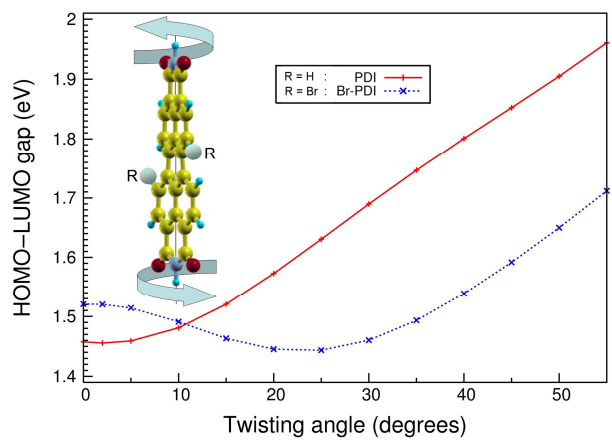

**Figure 4.** DFT HOMO–LUMO energy gap as a function of twisting angle about the major axis for PDI and Br-PDI.

All Br-PDI derived dye molecules studied in this paper have singlet HOMO and LUMO electronic states including Br-PDI itself. Although, the DFT gap values reflect infrared emission as a consequence of inability of including the *exact* exchange–correlation effects, TDDFT calculations produced accurate results which are presented in Table 1. The visible range absorbance spectra of these molecules are dominated by an intense HOMO-LUMO single electron transition as seen in Table 3. Consequently, our calculations suggest simple luminescence characteristics for Br-PDI derivatives involving solely the HOMO and LUMO, because the higher excited states either correspond to low-intensity bands or fall beyond visible region. We also noticed that the HOMO–LUMO charge distributions are similar to each other for Br-PDI, BrGly, and BrAsp. Therefore, calculated HOMO–LUMO 3D charge densities for BrGly are presented in Figure 3 as a representative of Br-PDI derived dye molecules. The involvement of Br $4p$ electrons shown in this figure supports the idea that the choice of halogen atom has a non-negligible effect on the energy level of HOMO and, hence, on the gap. The electrons seem to be localized around the center when the molecule is in its ground state more than the LUMO. When the molecule is excited to the LUMO, Br gives its HOMO charge to this charge transfer state which extends towards the ligands. We notice that the main difference in the charge distributions is that while the HOMO is aligned as if it is attracted by the two Br atoms, the LUMO looks like a flow of charge towards the tips where carboxyl groups are bound.

The optical absorption spectra for the remaining PDI derivatives reflect similar characteristics. Since we did not include spin flipping excitations, they all have a strong vertical transition from HOMO to LUMO excitation. TDDFT gives remarkably accurate results for the low lying single electron vertical excitation energies and oscillator strengths. These transitions can be considered as charge injecting states when they are considered as bound to a semiconductor surface. Yet, in this case, the excited spectra will be affected because of the dye-semiconductor interface[43,44]. Moreover, such an electron injection from the dye to the conduction band of the semiconductor might produce erroneous results because of the asymptotic long range behavior of charge transfer states when standard xc-functionals are used[45]. Because we do not study a charge transfer state of this kind, our gas phase TDDFT results presented in Table 2, 3 and 4 for all of the PDIs are expected to be accurate. In this sense, experimental verification of absorption properties are desirable.

**Table 3.** Calculated electronic spectra of dye sensitizers.

| | $\varepsilon$ [a] | $f$ [b] | $\Psi$ [c] |
|---|---|---|---|
| RuBpy | 2.62 | 0.0001 | $0.92|H \rightarrow L+1\rangle+0.92|H \rightarrow L+2\rangle$ |
| | 2.64 | 0.001 | $1.0|H \rightarrow L\rangle$ |
| | 2.85 | 0.004 | $0.75|H-1 \rightarrow L\rangle-0.36|H-2 \rightarrow L\rangle+0.35|H-2 \rightarrow L+1\rangle+0.35|H-1 \rightarrow L+2\rangle$ |
| | 2.85 | 0.004 | $0.75|H-2 \rightarrow L\rangle+0.36|H-1 \rightarrow L\rangle-0.35|H-2 \rightarrow L+2\rangle+0.35|H-1 \rightarrow L+1\rangle$ |
| | 2.96 | 0.11 | $-0.53|H-2 \rightarrow L+1\rangle+0.50|H-1 \rightarrow L\rangle-0.53|H-1 \rightarrow L+2\rangle$ |
| | 2.96 | 0.11 | $0.50|H-2 \rightarrow L\rangle+0.53|H-2 \rightarrow L+2\rangle-0.53|H-1 \rightarrow L+1\rangle$ |
| PDI | 2.45 | 0.68 | $1.0|H \rightarrow L\rangle$ |
| | 3.02 | 0.0002 | $0.97|H-1 \rightarrow L\rangle$ |
| | 3.61 | 0.025 | $-0.94|H-4 \rightarrow L\rangle-0.24|H \rightarrow L+4\rangle-0.20|H-8 \rightarrow L\rangle$ |
| Br-PDI | 2.39 | 0.49 | $1.0|H \rightarrow L\rangle$ |
| | 3.13 | 0.0005 | $0.87|H-4 \rightarrow L\rangle-0.43|H-2 \rightarrow L\rangle+0.20|H-3 \rightarrow L+1\rangle$ |
| | 3.18 | 0.0066 | $-0.94|H-1 \rightarrow L\rangle+0.23|H-3 \rightarrow L\rangle$ |
| BrGly | 2.37 | 0.55 | $1.0|H \rightarrow L\rangle$ |
| | 3.16 | 0.0058 | $0.96|H-1 \rightarrow L\rangle$ |
| | 3.21 | 0.0006 | $0.96|H-3 \rightarrow L\rangle$ |
| | 3.22 | 0.025 | $-0.93|H-4 \rightarrow L\rangle-0.26|H-2 \rightarrow L\rangle$ |
| | 3.30 | 0.070 | $0.92|H-2 \rightarrow L\rangle-0.27|H-4 \rightarrow L\rangle$ |
| BrAsp | 2.36 | 0.60 | $1.0|H \rightarrow L\rangle$ |
| | 3.13 | 0.0051 | $0.94|H-1 \rightarrow L\rangle$ |
| | 3.21 | 0.012 | $0.60|H-8 \rightarrow L\rangle-0.53|H-4 \rightarrow L\rangle+0.50|H-2 \rightarrow L\rangle$ |
| | 3.22 | 0.0015 | $-0.78|H-7 \rightarrow L\rangle+0.47|H-5 \rightarrow L\rangle+0.26|H-3 \rightarrow L\rangle$ |
| | 3.25 | 0.02 | $0.60|H-8 \rightarrow L\rangle-0.26|H-6 \rightarrow L\rangle-0.71|H-2 \rightarrow L\rangle$ |
| | 3.27 | 0.0015 | $-0.67|H-5 \rightarrow L\rangle+0.46|H-3 \rightarrow L\rangle-0.32|H-7 \rightarrow L\rangle-0.28|H-10 \rightarrow L\rangle+0.26|H-1 \rightarrow L\rangle$ |
| | 3.38 | 0.027 | $0.50|H-3 \rightarrow L\rangle-0.46|H-6 \rightarrow L\rangle+0.45|H-4 \rightarrow L\rangle+0.36|H-2 \rightarrow L\rangle+0.21|H-7 \rightarrow L\rangle-0.21|H \rightarrow L+1\rangle$ |

[a] Excitation energies in eV.
[b] Oscillator strengths
[c] Wavefunctions

In order to study the nature of the bonding between the halogen and the nearest neighbor carbon atoms we conducted Bader analysis based on a recent algorithm[46]. The results for partial charges on halogen atoms for some PDI-derived molecules are given in Table 4. The charge on F ions was found to be $-0.800e$ while it is $-0.194e$ and $-0.052e$ for Cl and Br ions in PDI molecules without any other ligand attached. These results suggest a covalent bonding. The amount of charge transfer implies increased polarization between C–F bonding in the case of F-PDI. In order to show the effect of ligands to Bader charges around halogen atoms we calculated the atomic charge for the cases of BrGly and BrAsp, too. The results show a lowering of the partial charge around Br ions by $-0.004e$ and $-0.003e$ for BrGly and BrAsp, respectively. This indicates no significant change in the Br–C bonding character for the PDI derived molecules.

**Table 4.** The results for the Bader analysis of halogen atoms.

| Molecule | Partial charges ($e$) | Dipole moment (e·Å) |
|---|---|---|
| F-PDI | –0.7996 | 1.425425 |
| Cl-PDI | –0.1937 | 0.254075 |
| Br-PDI | –0.0516 | 0.699630 |
| BrGly | –0.0475 | 0.729924 |
| BrAsp | –0.0484 | 0.702965 |

Neutral Ru atom has quintet ground state $^5F_5$ with an electronic configuration of $d^7s^1$. On the other hand, Ruthenium is considered in its Ru(II) form in the RuBpy complex so that it has a septet ground state $^7S_3$ with $d^5s^1$ configuration which allows to bind to six neighboring nitrogen ions to form octahedral complexes. Relaxed geometry and calculated molecular orbital energy spectra are presented in Figure 5. Our calculated ground state DFT molecular structure is in excellent agreement with the X-ray diffraction study of Rillema *et al*. For example, the value of 2.07 Å for Ru-N distance, in Table 1, compares well with the experimental bond length of 2.056 Å[47] while it was predicted to be 2.10 Å by Gorelsky *et al*. Similarly, the ordering and the composition of energy levels in Figure 5 matches the DFT results of Daul *et al* for the frontier molecular orbitals[25]. RuBpy molecule has singlet ground and spin restricted excited states with a HOMO–LUMO DFT gap of 1.85 eV which is predicted to be 2.64 eV by TDDFT. RuBpy exhibits two strong optically active transitions at 2.97 eV from HOMO-1,2 to LUMO+1,2 with almost equal intensities of *f*=0.11 in good agreement with both the theoretical result of Campbell *et al*[26] and the experimental result of Saes *et al*[23]. RuBpy has six optical transitions, four of which are bi-degenerate stemming from the degeneracy of HOMO-1 with HOMO-2 and LUMO+1 with LUMO+2 lying 0.08 eV below the HOMO and 0.13 eV above the LUMO, respectively. Therefore, visible range active excitations involve transitions which are mainly from Ru(II) $t_{2g}$ *d*–orbitals to (bpy) $\pi^*$–orbitals in nature. The availability of a number of low lying excited states increases the photon capture probabilities under visible light illumination.

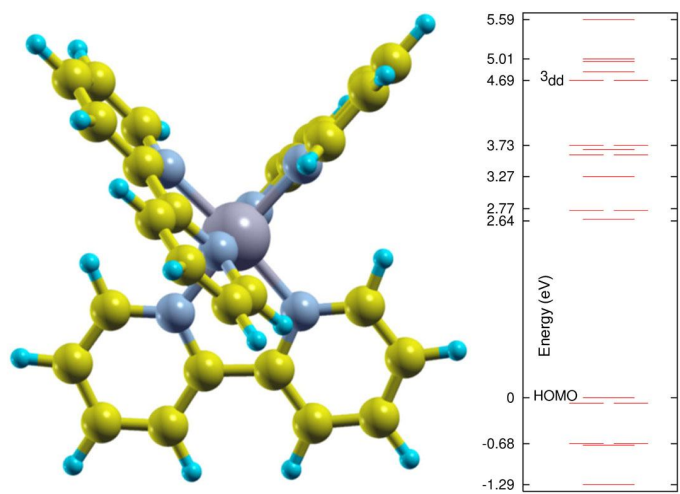

**Figure 5.** Calculated molecular structure and energy level spectrum of RuBpy dye molecule.

The role of Ru(II) is important in the catalytic activity of RuBpy dye molecule. Site-projected density of states analysis shows that the HOMO is populated by Ru $t_{2g}$-symmetry orbitals: $d_{xy}$, $d_{xz}$, and $d_{yz}$. So, this is a metal dominant state. On the other hand, the band decomposed 3D charge density distribution in Figure 6 for LUMO clearly shows that this is a metal-to-ligand charge transfer (MLCT) state. This can be referred as the lowest-lying $^1$MLCT state since the excitations to MLCT states do not undergo structural relaxation[48]. The metal localized triplet excited state, $^3dd$, can be assigned as the transitions to doubly degenerate LUMO+9 and LUMO+10 excitations. In Figure 7, we present the electronic charge distributions of these molecular orbitals which contribute to the $^3dd$ states which are totally populated by Ru antibonding $e_g$-orbitals and are 4.69 eV above the ground state. Fluorescence might not be simple upon absorption of light towards the violet because of the involvement of several transition paths from the excited states to the ground state. Thompson *et al*.[14] studied the decay mechanisms which involve (i) crossing down to the ground state via an $^3$MLCT state, (ii) direct deactivation to the HOMO, and (iii) ligand rearrangement forming $\pi$-bonded or three centered Ru–C–H agostic interaction to reform RuBpy.

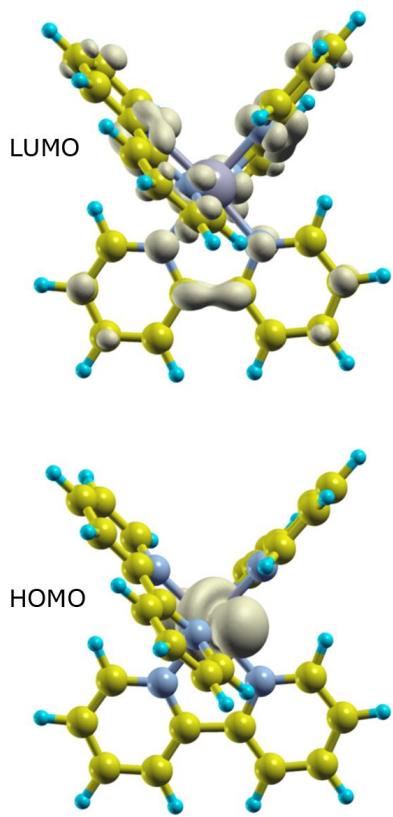

**Figure 6.** Calculated HOMO and LUMO charge density distributions of RuBpy dye molecule.

TDDFT results presented in Table 3 show the effect of the attached ligands on the light harvesting performances of Br-PDI complexes. The main contribution from the carboxylic groups is that they form a number of molecular orbitals below the HOMO which can be seen in Figure 2. This allows several optical transitions to occur basically from these MOs to the LUMO. The number of allowed valence excitations increases with the number of available MOs below the HOMO as a consequence of the attachment of more complex ligands. In the absence of such groups, i.e. in the case of PDI and Br-PDI, there are 2 and 3 visible range excitations, respectively. These excitations are dominated by one strong transition from the HOMO to the LUMO. The others exhibit very weak lines. However, when glycine or aspatine is attached to Br-PDI, 4 optical transitions are probable. We obtained 1 and 2 more near violet transitions with relatively higher intensities for BrGly and BrAsp, respectively. Therefore, even though carboxylic groups do not have a significant effect on the HOMO-LUMO excitation which is assigned as the gap value, they rather play a considerable role in enhancing the photon capture capability of these chromophores. TDDFT calculations suggest that BrAsp performs better light harvesting compared to BrGly both because of the larger number of visibly active excitations and because of the relatively stronger intensities.

The dyes are known to be anchored to the semiconductor surface by the carboxyl groups. It has been reported that as the number of the carboxyl groups increase in the dye sensitizers, the electron transfer efficiencies increased due to their better anchoring to the surface[4]. It is highly likely that, not the HOMO-LUMO levels, but the effective anchoring on the surface of the oxide is of primary importance in the enhanced charge injection yields observed with BrAsp in comparison to BrGly dyes.

As a result of the theoretical calculations, it was concluded that the HOMO–LUMO gap does not depend on the carboxylic acid groups but on the halogen atom. In the presence of more carboxylic acid groups (as in BrAsp relative to BrGly) the rate enhancement during the photocatalytic reaction probably lies in the fact that the dye sensitizers could be anchored on $TiO_2$ at more contact points therefore the electron transfer efficiencies were increased.

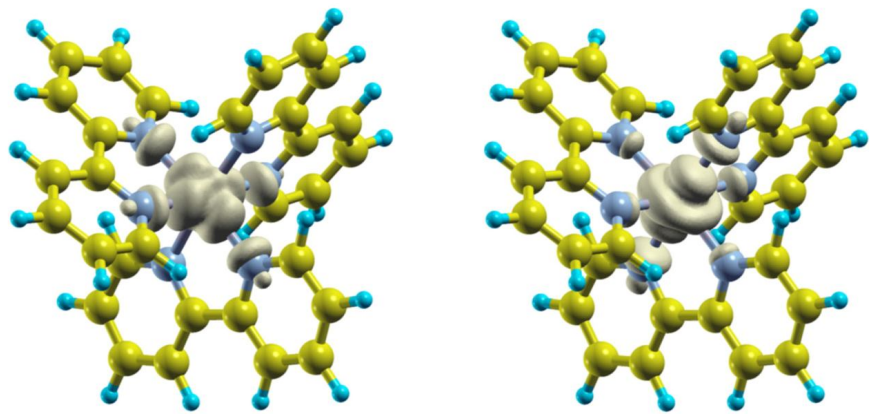

**Figure 7.** 3D charge density plots of doubly degenerate unoccoupied molecular orbitals which contribute to the $^3dd$ states of RuBpy.

CONCLUSIONS

Theoretical calculations show that the choice of halogen atom is crucial for light absorbing activity of the PDI dye molecules. Electronically, the HOMO level has contribution from the *p*-electrons of the two halogen atoms. In addition, both halogen atoms incorporate repulsive stress to the dye molecule from the sides where they bind. This forces the whole structure to be twisted in order to account for relaxation towards the lowest energy geometry. The larger is the atomic radius of the halogen, the larger the twisting of the molecule is. Yet, this is not the only factor which affects the twist angle. Our results suggest that the choice of carboxylic groups has a relatively less but non-negligible effect on this reconstruction. The atomic arrangement of the molecule has a direct influence on the LUMO level, since it is populated by the electrons in the perylene part of the molecule. Therefore, both the choice of halogen species and the lowest energy geometric structure of the PDI-derived dyes are responsible for the formation of the HOMO–LUMO gap. In other words, it is possible to tune the gap and therefore the light harvesting spectrum of the dye molecule by choosing the halogen species and attaching different carboxylic functional groups.

The light absorption characteristics of PDI-derived and RuBpy dyes show some differences. In the case of PDI dyes photon is absorbed by the ground state which extends through perylene part including the *p*-electrons of the halogen atom. However, for RuBpy, this absorption is fulfilled solely by the $t_{2g}$-symmetry *d*-orbitals of the ruthenium ion. We expect light harvesting RuBpy to perform relatively better in visible region because of the involvement of more than one excited state. On the other hand, the charge distributions of LUMO states for both types of dye complexes extend to tips. These states allow electron transfer to occur from ligands to possible contacts. Therefore, not only the electronic structure but also the types of ligands play an important role in the charge injection efficiency of the dyes. Our calculations suggest that Cl-, Br-, and F-PDI derivatives might be promising alternatives to RuBpy dye sensitizers.

While the results exhibit small differences between the HOMO–LUMO gaps of the Br-PDI derivatives, the experimental results indicate enhanced activity when more carboxylic acid groups are present[31]. We interpret the results as new carboxyl groups added, the number of carbon chain branches grow like a tree leading to increased number of probable contacts with the host surface, thus enhancing the electron transfer efficiencies. Our results suggested that, the photon absorption performance of PDIs can be enhanced by attaching more complex ligands which form new sub-MOs in the vicinity of the HOMO as well as the choice of the halogen species.

ACKNOWLEDGMENT

Financial support from TUBITAK, The Scientific and Technical Research Council of Turkey, under research grant no TBAG-HD/38 (105T051) is gratefully acknowledged. Further support was provided by Middle East Technical University BAP-2004-07-02-00-100 and YUUP-BAP 2004-08-11-06 Projects.